\documentstyle[preprint,aps,epsf,floats,tighten]{revtex} 

\begin{document}

\preprint{\tighten \vbox{\hbox{   }
 \hbox{} \hbox{} \hbox{} \hbox{} \hbox{} } }

\title{Higher twist corrections to the sum rule for semileptonic $B$ decay }

\author{Changhao Jin}

\address{School of Physics,
University of Melbourne\\Victoria 3010, Australia}

\maketitle

{\tighten
\begin{abstract}%
The sum rule for charmless inclusive semileptonic $B$-meson decays allows
a theoretically clean and experimentally efficient determination of
$|V_{ub}|$. The leading twist contribution to the sum rule is known
in QCD. We compute higher twist corrections to the sum rule using the heavy 
quark effective theory.
\end{abstract}
}%end tighten

\newpage

A new method to determine the Cabibbo-Kobayashi-Maskawa (CKM) matrix 
element $|V_{ub}|$ has been proposed \cite{sum} that takes advantage of
the sum rule for charmless inclusive semileptonic $B$-meson decays
$\bar B\to X_u\ell\bar{\nu_\ell}$ ($\ell=e$ or $\mu$). The sum rule 
establishes a clean relationship between $|V_{ub}|$ and the observable 
\begin{equation}
S\equiv\int_0^1 d\xi_u\, \frac{1}{\xi_u^5}
\frac{d\Gamma}{d\xi_u}(\bar B\to X_u\ell\bar{\nu_\ell}) 
\label{eq:integral}
\end{equation}
with the kinematic variable $\xi_u=(q^0+|{\bf q}|)/M_B$
in the $B$-meson rest frame, where $q$ is the momentum transfer to the lepton 
pair and $M_B$ denotes the $B$ meson mass.
Moreover, this method of determining $|V_{ub}|$ has experimental virtue too.
The kinematic variable $\xi_u$ is the most efficient discriminator between 
$\bar B\to X_u\ell\bar{\nu_\ell}$ signal and $\bar B\to X_c\ell\bar{\nu_\ell}$
background.
A majority of $\bar B\to X_u\ell\bar{\nu_\ell}$ events have a value of 
$\xi_u$ beyond the limit allowed for $\bar B\to X_c\ell\bar{\nu_\ell}$ 
decays with charm in the final state, $\xi_u > 1-M_D/M_B = 0.65$ with $M_D$ 
being the $D$ meson mass.
Therefore, only a small extrapolation is needed to obtain $S$.

The charmless inclusive semileptonic decay of the $B$ meson is a light-cone
dominated process. The light-cone expansion allows a rigorous and systematic
ordering of nonperturbative QCD effects, providing an effective technique
for a separation and classification of higher twist (HT) effects \cite{LC}. 
The leading term in this 
expansion gives the leading twist contribution. HT contributions are 
contained in the light-cone expansion beyond the leading order.
The sum rule at the leading twist order measures the bottomness carried by a
$B$ meson. There are no perturbative QCD corrections to the sum rule.  
Thus the primary
hadronic uncertainty and the potential uncertainty of perturbative QCD are
eliminated, dramatically reducing the theoretical error on $|V_{ub}|$.
This inclusive method is to be contrasted with the determination of $|V_{ub}|$
from the charmless inclusive semileptonic branching fraction of $B$ mesons 
where the calculation of the total semileptonic decay rate is model dependent 
or assumes quark-hadron duality,
there are uncertainties due to perturbative QCD corrections and, in addition,
a larger extrapolation is necessary to extract the total rate if the 
kinematic cut on a certain observable, such as the charged-lepton energy
or the invariant mass of the lepton pair, 
is applied for the suppression of $b\to c$ background.
   
Only uncertainties due to HT effects remain in the sum rule.
Including the HT contribution $\Delta HT$, the sum rule reads

\begin{equation}
S\equiv\int_0^1 d\xi_u\, \frac{1}{\xi_u^5}
\frac{d\Gamma}{d\xi_u}(\bar B\to X_u\ell\bar{\nu_\ell}) 
= |V_{ub}|^2\frac{G_F^2M_B^5}{192\pi^3}(1+\Delta HT)\, .
\label{eq:sumrule}
\end{equation}
Although HT
contributions are expected to be suppressed by powers of 
$\Lambda_{\rm QCD}^2/M_B^2$ ($\Lambda_{\rm QCD}$ being the QCD scale),
a quantitative estimate of them is indispensable for a complete understanding 
of remaining theoretical uncertainties in this determination of
$|V_{ub}|$. In this paper,
we investigate HT effects on the sum rule for charmless
inclusive semileptonic $B$ decays
using the heavy quark effective theory (HQET) \cite{isgur,eichten,georgi}.

Charmless inclusive semileptonic decays of the $B$ meson are induced by the 
weak interactions. 
The differential decay rate to lowest order in the weak interactions is 
\begin{equation}
d\Gamma= \frac{G_{F}^2\left|V_{ub}\right|^2}{(2\pi)^5E}L^{\mu\nu}
W_{\mu\nu}\frac{d^3k_\ell}{2E_{\ell}}\frac{d^3k_{\nu}}{2E_{\nu}} \, .
\label{eq:kga}
\end{equation}
Here $E$ ($P$), $E_\ell$ ($k_\ell$), and $E_\nu$ ($k_\nu$) denote the energies 
(four-momentums) of the $B$ meson, the charged lepton, and the antineutrino, 
respectively. 
The leptonic tensor for the lepton pair is completely determined by the
standard electroweak theory since leptons do not have strong interactions: 
\begin{equation}
L^{\mu\nu}= 2(k^{\mu}_{\ell}k^{\nu}_{\nu}+k^{\mu}_{\nu}k^{\nu}_{\ell}-
g^{\mu\nu}k_{\ell}\cdot k_{\nu}+i\varepsilon^{\mu\nu}\hspace{0.06cm}_{\alpha\beta}
k^{\alpha}_{\ell}k^{\beta}_{\nu}).
\label{eq:klepton}
\end{equation}
The hadronic tensor incorporates all nonperturbative QCD physics for the 
inclusive semileptonic $B$ decay. It is 
summed over all hadronic final states and can be expressed in terms of a 
current commutator taken between the $B$ meson states:
\begin{equation}
W_{\mu\nu}= -\frac{1}{2\pi}\int d^4y e^{iq\cdot y}
\langle B\left|[j_{\mu}(y),j^{\dagger}_{\nu}(0)]\right|B\rangle ,
\label{eq:comm2}
\end{equation}
where $j_{\mu}(y) = \bar{u}(y)\gamma_{\mu}(1-\gamma_5)b(y)$ is 
the charged weak current for the $b\to u$ transition. We adopt
a covariant normalization for one-particle states, i.e.,
$\langle B(P)|B(P^\prime)\rangle = (2\pi)^3 2P^0\delta^{(3)}({\bf P}-
{\bf P}^\prime)$. 
 
The most general hadronic tensor form that can be constructed is a 
linear combination of $P_\mu P_\nu,\,
P_\mu q_\nu,\, q_\mu P_\nu,\, q_\mu q_\nu,\, 
\varepsilon_{\mu\nu\alpha\beta}P^\alpha q^\beta$, and $g_{\mu\nu}$, 
with coefficients being scalar functions $W_a(\eta,\, q^2)$ of the two 
independent Lorentz invariants,
$\nu\equiv q\cdot P/M_B$ and $q^2$.  However,   
the combination $P_\mu q_\nu-q_\mu P_\nu$ does not contribute
since $L^{\mu\nu}(P_\mu q_\nu-q_\mu P_\nu)=0$.  Thus the hadronic tensor 
must take the form

\begin{equation}
W_{\mu\nu} = -g_{\mu\nu}W_1 + \frac{P_{\mu}P_{\nu}}{M_B^2} W_2 
 -i\varepsilon_{\mu\nu\alpha\beta} \frac{P^{\alpha}q^{\beta}}{M_B^2}W_3
         + \frac{q_{\mu}q_{\nu}}{M_B^2} W_4 
 + \frac{P_{\mu}q_{\nu}+q_{\mu}P_{\nu}}{M_B^2} W_5 \, .
\label{eq:exp2}
\end{equation}
Equation (\ref{eq:comm2}) shows that $W_{\mu\nu}^{\ast}=W_{\nu\mu}$,
so $W_a$, $a=1,\ldots, 5$ are real.
The interesting physics describing the hadron structure and the
strong interactions is wrapped up in the five dimensionless real structure 
functions $W_a(\nu,\, q^2)$, $a=1,\ldots, 5$ for the unpolarized processes.

In the following we will neglect the masses of the charged lepton and
the $u$-quark. From Eqs.~(\ref{eq:kga}) and (\ref{eq:exp2}), we obtain
the double differential decay rate for $\bar B\to X_u\ell\bar{\nu_\ell}$
in the rest frame of the $B$ meson

\begin{equation}
\frac{d^2\Gamma}{d\xi_u dq^2} = \frac{G_{F}^2\left|V_{ub}\right|^2}{48\pi^3 M_B}
\frac{|{\bf q}|^2}{\xi_u}(W_1 3q^2+W_2 |{\bf q}|^2),
\label{eq:double}
\end{equation}
where 

\begin{equation}
|{\bf q}|=\frac{1}{2}M_B\xi_u(1-\frac{q^2}{M_B^2\xi_u^2}).
\end{equation}
By integrating Eq.~(\ref{eq:double}) over $q^2$, one gets the decay 
distribution of the kinematic variable $\xi_u$

\begin{equation}
\frac{d\Gamma}{d\xi_u} = \int_0^{M_B^2\xi_u^2}dq^2 \frac{d^2\Gamma}{d\xi_u dq^2}.
\label{eq:gaxiu}
\end{equation}

Computing the current commutators one obtains from Eq.~(\ref{eq:comm2})

\begin{equation}
W_{\mu\nu}
 = -\frac{1}{\pi}(S_{\mu\alpha\nu\beta} -i\varepsilon_{\mu\alpha\nu\beta})
\int d^4 y  e^{iq\cdot y}\left[ \partial^{\alpha}\Delta_u (y) \right] 
\langle B\left|\bar{b}(0)\gamma^{\beta}U(0,y)b(y)\right|B\rangle \, ,
\label{eq:comm3}
\end{equation}
where $S_{\mu\alpha\nu\beta} = g_{\mu\alpha}g_{\nu\beta} + g_{\mu\beta}
 g_{\nu\alpha} - g_{\mu\nu}g_{\alpha\beta}$.
In the above we have used

\begin{eqnarray}
\{u(x), \bar u(y)\} = i(\gamma\cdot \partial) i\Delta_{u}(x-y) U(x,y)
\end{eqnarray}
with the Wilson link

\begin{eqnarray}
U(x,y) &=& {\cal P} \mbox{exp}[ig_s \int^x_y dz^\mu A_\mu(z)] ,
\label{eq:link}\\
\Delta_{u}(y) &=& -{i\over (2\pi)^3} \int d^4k e^{-ik\cdot y}\varepsilon(k^0)
\delta(k^2),
\end{eqnarray}
where $A^\mu$ is the background gluon field and $\varepsilon(x)$ satisfies
$\varepsilon(|x|)=1$ and $\varepsilon(-|x|)= -1$.

The matrix element $\langle B|\bar b(0) \gamma^\beta U(0,y)b(y)|B\rangle$ 
is the basic 
building block of the description of inclusive $B$ decays in QCD.
In general one can decompose it in the following form:

\begin{eqnarray}
\langle B|\bar b(0) \gamma^\beta U(0,y)b(y)|B\rangle = 
2[P^\beta F(y^2, y\cdot P) + 
y^\beta G(y^2,y\cdot P)],
\end{eqnarray}
where $F(y^2, y\cdot P)$ and $G(y^2, y\cdot P)$ are functions of the two
independent Lorentz scalars, $y^2$ and $y\cdot P$.
The dominant part of the integrand in the hadronic tensor
(\ref{eq:comm3}) stems from the space-time region 
near the light cone, with the deviation from the light cone being
of the order of the inverse large momentum 
$y^2\sim 1/q^{2}\sim 1/M_B^{2}\to 0$ \cite{LC}. 
The light-cone expansion of the functions
$F(y^2, y\cdot P)$ and $G(y^2, y\cdot P)$ in powers of $y^2$ leads to

\begin{eqnarray}
&&\langle B|\bar b(0) \gamma^\beta U(0,y)b(y)|B\rangle = 
2\left[ P^\beta \sum_{n=0}^\infty 
(y^2)^n {\cal F}^{(2n+2)}(y\cdot P) + 
y^\beta \sum_{n=0}^\infty (y^2)^n {\cal G}^{(2n+4)}(y\cdot P)\right]\nonumber\\
&&= 2\left\{ P^\beta \left[ {\cal F}^{(2)}(y\cdot P) + y^2 {\cal F}^{(4)}(y\cdot P) 
+ \cdots \right]+
y^\beta \left[ {\cal G}^{(4)}(y\cdot P) + y^2 {\cal G}^{(6)}(y\cdot P) 
+ \cdots \right]\right\} .
\label{eq:lightexp}
\end{eqnarray}
The coefficients ${\cal F}^{(2n+2)}(y\cdot P)$ and
${\cal G}^{(2n+4)}(y\cdot P)$ in the light-cone expansion can be classified
by twist.
Following the notion of twist introduced by Jaffe and Ji \cite{jaffe},
${\cal F}^{(2n+2)}(y\cdot P)$ has twist $2n+2$ and 
${\cal G}^{(2n+4)}(y\cdot P)$ has twist
$2n+4$, as from dimension analysis we know that the contribution of the 
former is suppressed by $(\Lambda_{\rm QCD}/M_B)^{2n}$, 
and the contribution of the latter
is suppressed by $(\Lambda_{\rm QCD}/M_B)^{2n+2}$. We will further discuss 
the non-local light-cone expansion of matrix elements below.

The twist decomposition for the decay rate thus takes the form

\begin{equation}
d\Gamma = \sum_{n=0}^\infty d\Gamma^{(2n+2)} ,
\end{equation}
where 

\begin{equation}
d\Gamma^{(2n+2)} = \frac{G_F^2\left|V_{ub}\right|^2}{(2\pi)^5E}L^{\mu\nu}
W_{\mu\nu}^{(2n+2)}\frac{d^3k_\ell}{2E_\ell}\frac{d^3k_\nu}{2E_\nu} 
\label{eq:gatwist}
\end{equation}
is the twist-$2n+2$ contribution to the decay rate with

\begin{eqnarray}
W_{\mu\nu}^{(2n+2)} &=& -g_{\mu\nu}W_1^{(2n+2)} + 
\frac{P_\mu P_\nu}{M_B^2} W_2^{(2n+2)} 
-i\varepsilon_{\mu\nu\alpha\beta} \frac{P^\alpha q^\beta}{M_B^2}W_3^{(2n+2)}
\nonumber \\
&& + \frac{q_\mu q_\nu}{M_B^2} W_4^{(2n+2)} 
 + \frac{P_\mu q_\nu+q_\mu P_\nu}{M_B^2} W_5^{(2n+2)}  \label{eq:twdef}\\
&=& -\frac{2}{\pi}(S_{\mu\alpha\nu\beta} -i\varepsilon_{\mu\alpha\nu\beta})
\int d^4 y  e^{iq\cdot y}\left[ \partial^\alpha \Delta_u (y) \right] 
\nonumber \\
&& \times \left[ P^\beta (y^2)^n {\cal F}^{(2n+2)}(y\cdot P) + 
y^\beta (y^2)^{n-1} {\cal G}^{(2n+2)}(y\cdot P)\right] \, .
\label{eq:twten}
\end{eqnarray}

The leading twist contribution to the sum rule 
(\ref{eq:sumrule}) results from ${\cal F}^{(2)}(y\cdot P)$ of twist 2 and 
is known in QCD to be \cite{sum}

\begin{equation}
\int_0^1 d\xi_u\, \frac{1}{\xi_u^5}
\frac{d\Gamma^{(2)}}{d\xi_u}(\bar B\to X_u\ell\bar{\nu_\ell}) 
= |V_{ub}|^2\frac{G_F^2M_B^5}{192\pi^3},
\label{eq:leading}
\end{equation}
which is a consequence of the conservation of the $b$-quark vector 
current by the strong interactions.
The next-to-leading twist contribution to the sum rule arises from  
${\cal F}^{(4)}(y\cdot P)$ and ${\cal G}^{(4)}(y\cdot P)$ of twist 4. It
can be obtained by integrating 
Eq.~(\ref{eq:gaxiu}) over $\xi_u$ with the two relevant structure 
functions $W_1^{(4)}(\nu, q^2)$ and $W_2^{(4)}(\nu, q^2)$ of twist 4. 

We use the operator product expansion and the heavy quark effective theory to 
compute the twist-4 structure functions.
The Wilson link is a gauge dependent operator.
It is convenient to use 
the Fock-Schwinger gauge such that $U(0,y)$ is unity.
Since the $b$ quark inside the $B$ meson behaves as almost free due to its 
large mass, relative to which its binding to the light constituents is weak,
one can extract the large space-time dependence

\begin{equation}
b(y)= e^{-im_bv\cdot y}b_v(y) ,
\label{eq:scale}
\end{equation}
where $m_b$ is the $b$-quark mass and 
$v=P/M_B$ is the four-velocity of the $B$ meson. 
This factorization makes clear 
why the large scale in matrix elements does not affect the relative size of 
terms in the light-cone expansion (\ref{eq:lightexp}). The large scale hidden
in matrix elements of $b$-quark operators is contained in an overall factor
$e^{-im_bv\cdot y}$, so reduced matrix elements of the operators containing
the rescaled 
operator $b_v$ involve only momenta of order $\Lambda_{\rm QCD}$, which
determine the relative size of terms in the light-cone expansion
(\ref{eq:lightexp}), i.e., schematically

\begin{equation}
\langle B|\bar b(0) \gamma^\beta b(y)|B\rangle = e^{-im_bv\cdot y}
\langle B|\bar b_v(0) \gamma^\beta b_v(y)|B\rangle 
\sim e^{-im_bv\cdot y}\sum_{n=0}^\infty 
\left(\frac{\Lambda^2_{\rm QCD}}{M_B^2}\right)^n.
\end{equation}

The rescaled operator for a free $b$-quark
no longer depends on the space-time, so $b(y) = e^{-im_bv\cdot y}b(0)$. 
In this case all the coefficients ${\cal F}^{(2n+2)}(y\cdot P)$ and
${\cal G}^{(2n+4)}(y\cdot P)$ in the light-cone expansion (\ref{eq:lightexp}) 
vanish except that ${\cal F}^{(2)}(y\cdot P) = e^{-im_bv\cdot y}$, because
the conservation of the $b$-quark vector current implies that
$\langle B|\bar b(0) \gamma^\beta b(0)|B\rangle = 2P^\beta$. The 
leading-twist sum rule (\ref{eq:leading}) is consistently reproduced in the
free quark decay $b\to u\ell\bar{\nu_\ell}$. The conserved vector current
$\bar b\gamma^\beta b$ is not renormalized by the strong interactions. 
This explains why there are no perturbative QCD corrections to the sum rule.

A Taylor expansion of the field in a gauge-covariant form relates the 
bilocal and local operators. This leads to an operator product expansion

\begin{equation}
\bar b(0)\gamma^\beta b(y)= e^{-im_bv\cdot y}\bar b_v(0)\gamma^\beta b_v(y)
= e^{-im_bv\cdot y}\sum_{n=0}^\infty 
\frac{(-i)^n}{n!}
y_{\mu_1}\cdots y_{\mu_n}\bar b_v(0)\gamma^\beta 
k^{\{\mu_1}\cdots k^{\mu_n\} }b_v(0) ,
\label{eq:moper}
\end{equation}
where $k_\mu=iD_\mu=i(\partial_\mu-ig_sA_\mu)$ and the symbol
$\{\cdots\}$ means symmetrization with respect to the enclosed indices.
Because of the weak dependence of the rescaled operator $b_v(y)$ on $y$, 
we attempt to estimate the matrix element of the bilocal operator sandwiched
between the $B$ meson states with the truncated $y$-expansion in 
Eq.~(\ref{eq:moper}). To obtain a twist-4 accuracy it suffices to keep 
only the first three terms

\begin{eqnarray}
\langle B|\bar b(0)\gamma^\beta b(y)|B\rangle = &&e^{-im_bv\cdot y}
\Bigg [\langle B|\bar b_v(0)\gamma^\beta b_v(0)|B\rangle  \nonumber \\
&& +(-i)y_\mu\langle B|\bar b_v(0)\gamma^\beta iD^\mu b_v(0)|B\rangle  
\nonumber\\
&& +\frac{(-i)^2}{2}y_\mu y_\nu
\langle B|\bar b_v(0)\gamma^\beta iD^{\{\mu}iD^{\nu\}}b_v(0)|B\rangle\Bigg ].
\label{eq:mtrun}
\end{eqnarray}

In the heavy quark effective theory 
the QCD $b$-quark field $b(y)$ is related to 
its HQET counterpart $h(y)$ by means of an expansion in
powers of $1/m_b$:

\begin{equation}
b(y)=e^{-im_bv\cdot y}\left[ 1+\frac{i/ \mkern -12mu D}{2m_b}+
O\left( \frac{\Lambda^2_{\rm QCD}}{m_b^{2}}\right) \right]h(y) .
\label{eq:mexp}
\end{equation}
The effective Lagrangian takes the form

\begin{equation}
{\cal L}_{\rm HQET}= \bar hiv\cdot Dh+\bar h\frac{(iD)^2}{2m_b}
h
 +\bar h\frac{g_sG_{\mu\nu}\sigma^{\mu\nu}}{4m_b}
h+O\left(\frac{1}{m_b^2}\right) ,
\label{eq:mLag}
\end{equation}
where $g_sG^{\mu\nu}=i[D^\mu, D^\nu]$ is the gluon field-strength
tensor. At the level of accuracy of the present discussion we take into 
account only the leading, $1/m_b$ correction to the heavy quark limit
$m_b\to \infty$.
Relating the matrix elements of the local operators in full QCD
in Eq.~(\ref{eq:mtrun}) to those in HQET, it follows that 

\begin{eqnarray}
\langle B|\bar b(0)\gamma^\beta b(y)|B\rangle = &&2e^{-im_bv\cdot y}
\Bigg\{ P^\beta \left[1-y\cdot P i\frac{5}{3}\frac{m_b}{M_B}E_b-
(y\cdot P)^2 \frac{1}{3}\frac{m_b^2}{M_B^2}K_b+
y^2\frac{1}{3}m_b^2K_b\right]  \nonumber \\
&&+y^\beta i\frac{2}{3}m_bM_BE_b \Bigg\} ,
\label{eq:mexp}
\end{eqnarray}
where $E_b=K_b+G_b$ and $K_b$ and $G_b$ are the dimensionless HQET parameters 
of order $(\Lambda_{\rm QCD}/m_b)^2$, which are often referred to by the
alternate names $\lambda_1= -2m_b^2K_b$ and $\lambda_2= -2m_b^2G_b/3$,
defined as

\begin{eqnarray}
\lambda_1 &=& \frac{1}{2M_B}\langle B|\bar h
(iD)^2 h|B\rangle , \label{eq:mK} \\
\lambda_2 &=& \frac{1}{12M_B}\langle B|\bar h g_sG_{\mu\nu}
\sigma^{\mu\nu} h|B\rangle . 
\label{eq:def}
\end{eqnarray}

Comparing Eq.~(\ref{eq:mexp}) with Eq.~(\ref{eq:lightexp}) yields

\begin{eqnarray}
{\cal F}^{(4)}(y\cdot P) &=& \frac{1}{3}m_b^2K_b e^{-im_bv\cdot y},
\label{eq:f4}\\
{\cal G}^{(4)}(y\cdot P) &=& i\frac{2}{3}m_bM_BE_b e^{-im_bv\cdot y}.
\label{eq:g4}
\end{eqnarray}
We observe that the coefficients ${\cal F}^{(4)}(y\cdot P)$ and 
${\cal G}^{(4)}(y\cdot P)$ of the light-cone expansion (\ref{eq:lightexp}) are 
indeed of order $\Lambda_{\rm QCD}^2$ as expected.
Substituting Eqs.~(\ref{eq:f4}) and (\ref{eq:g4}) in Eq.~(\ref{eq:twten}) and
integrating by parts, we arrive at

\begin{eqnarray}
W^{(4)}_{\mu\nu} = && \frac{16m_b}{3M_B}\Bigg\{ -g_{\mu\nu}M_B^2
\Big [\frac{1}{4}m_b
(m_b-\nu)K_bX \nonumber \\
&& +E_b\varepsilon (q^0-m_bv^0)\delta(q^2-2m_b\nu+m_b^2) \nonumber \\
&& +m_b(m_b-\nu)K_b\varepsilon (q^0-m_bv^0)\delta^\prime(q^2-2m_b\nu+m_b^2)
\nonumber \\
&& +(q^2-2m_b\nu+m_b^2)E_b\varepsilon (q^0-m_bv^0)\delta^\prime(q^2-2m_b\nu+m_b^2)\Big ]
\nonumber \\
&& +P_\mu P_\nu m_b^2 \left[\frac{1}{2}K_bX+2(K_b+E_b)
\varepsilon (q^0-m_bv^0)\delta^\prime(q^2-2m_b\nu+m_b^2)\right] \nonumber \\
&& +i\varepsilon_{\mu\nu\alpha\beta}P^\alpha q^\beta m_bM_BK_b
\left[\frac{1}{4}X+
\varepsilon (q^0-m_bv^0)\delta^\prime(q^2-2m_b\nu+m_b^2)\right] \nonumber \\
&& +q_\mu q_\nu 2M_B^2E_b\varepsilon (q^0-m_bv^0)\delta^\prime(q^2-2m_b\nu+m_b^2) 
\nonumber\\
&& +(P_\mu q_\nu+q_\mu P_\nu) m_bM_B\Big [-\frac{1}{4}K_bX-
K_b\varepsilon (q^0-m_bv^0)\delta^\prime(q^2-2m_b\nu+m_b^2) \nonumber \\
&& -2E_b\varepsilon (q^0-m_bv^0)\delta^\prime(q^2-2m_b\nu+m_b^2)\Big ]\Bigg\},
\label{eq:t4W}
\end{eqnarray}
where $\delta^\prime(x) = \frac{d}{dx}\delta(x)$ and

\begin{equation}
X = \frac{\partial^2}{\partial q^\mu\partial q_\mu}
[\varepsilon (q^0-m_bv^0)\delta (q^2-2m_b\nu+m_b^2)].
\end{equation}
Comparing Eq.~(\ref{eq:t4W}) with Eq.~(\ref{eq:twdef}), we find

\begin{eqnarray}
W_1^{(4)}(\nu, q^2) =&& \frac{16}{3}m_bM_B\Bigg\{ \frac{1}{4}m_b(m_b-\nu)K_bX+
E_b\varepsilon (q^0-m_bv^0)\delta (q^2-2m_b\nu+m_b^2) \nonumber \\
&& +[m_b(m_b-\nu)K_b+(q^2-2m_b\nu+m_b^2)E_b] \nonumber \\
&& \times\varepsilon (q^0-m_bv^0)\delta^\prime (q^2-2m_b\nu+m_b^2)\Bigg\}, 
\label{eq:w1t4}\\
W_2^{(4)}(\nu, q^2) =&& \frac{16}{3}m_b^3M_B \left[\frac{1}{2}K_bX+
2(K_b+E_b)\varepsilon (q^0-m_bv^0)\delta^\prime (q^2-2m_b\nu+m_b^2)\right],
\label{eq:w2t4}\\
W_3^{(4)}(\nu, q^2) =&& -\frac{16}{3}m_b^2M_B^2K_b \left[\frac{1}{4}X+
\varepsilon (q^0-m_bv^0)\delta^\prime (q^2-2m_b\nu+m_b^2)\right],
\label{eq:w3t4}\\
W_4^{(4)}(\nu, q^2) =&& \frac{32}{3}m_bM_B^3 E_b
\varepsilon (q^0-m_bv^0)\delta^\prime (q^2-2m_b\nu+m_b^2),
\label{eq:w4t4}\\
W_5^{(4)}(\nu, q^2) =&& -\frac{16}{3}m_b^2M_B^2 \left[\frac{1}{4}K_bX+
(K_b+2E_b)\varepsilon (q^0-m_bv^0)\delta^\prime (q^2-2m_b\nu+m_b^2)\right].
\label{eq:w5t4}
\end{eqnarray}

The twist-4 contribution to the sum rule can be obtained from 
Eqs.~(\ref{eq:gaxiu}),
(\ref{eq:double}), (\ref{eq:w1t4}) and (\ref{eq:w2t4}). The result is

\begin{eqnarray}
\int_0^1d\xi_u\frac{1}{\xi_u^5}
\frac{d\Gamma^{(4)}}{d\xi_u}(\bar B\to X_u\ell\bar{\nu_\ell}) =&& 
|V_{ub}|^2\frac{G_F^2M_B^5}{192\pi^3}
\Bigg [\frac{304}{45}K_b+\frac{76}{45}E_b
+2\frac{m_b^2}{M_B^2} K_b 
-\frac{68}{9}\frac{m_b^3}{M_B^3} K_b
 \nonumber \\
&& -\frac{80}{9}\frac{m_b^3}{M_B^3} E_b
-\frac{26}{3}\frac{m_b^4}{M_B^4} K_b
+\frac{28}{3}\frac{m_b^4}{M_B^4} E_b
+\frac{112}{15}\frac{m_b^5}{M_B^5} K_b
 \nonumber \\
&& -\frac{32}{15}\frac{m_b^5}{M_B^5} E_b\Bigg ].
\label{eq:tw4sum}
\end{eqnarray}
This can serve as an estimate of HT contributions to
the sum rule (\ref{eq:sumrule}).

For the numerical analysis, we need to know the values for the 
parameters involved. The HQET parameter $\lambda_2$ can be extracted from
the $B^\ast-B$ mass
splitting: $\lambda_2 = (M^2_{B^\ast}-M^2_B)/4 \simeq 0.12$ GeV$^2$,
while $\lambda_1$ and $m_b$ are less determined. 
For the purpose of estimation, we take 
$\lambda_1 = -0.5$ GeV$^2$ \cite{ball,latt}
and $m_b = 4.9$ GeV \cite{PDG}. 
From Eq.~(\ref{eq:tw4sum}), the HT correction 
to the sum rule (\ref{eq:sumrule}) is then
estimated to be $\Delta HT = 0.012$. This quantitative study shows that HT
corrections are at the expected level of 
$\sim \Lambda_{\rm QCD}^2/M_B^2$.

In summary, we have elaborated a quantitative way of estimating HT 
contributions in inclusive $B$ decays, which is based on the heavy quark 
effective theory. As an application we have calculated the twist-4 
correction to the sum rule (\ref{eq:sumrule}) for charmless inclusive
semileptonic $B$ decays. 
Using the sum rule,
$|V_{ub}|$ can be determined from a measurement of the weighted integral $S$.
The error on $|V_{ub}|$
due to HT corrections to the sum rule is estimated to be $1\%$.
Combining theoretical cleanliness and experimental efficiency, together with
the better understanding of remaining theoretical uncertainties,
the sum rule holds high promise of
a precise and model-independent determination of $|V_{ub}|$.

\acknowledgments
Stimulating discussions with Xiao-Gang He and Berthold Stech are gratefully
acknowledged.
This work was supported by the Australian Research Council.

{\tighten

} %end tighten (references & figure captions)


\begin{references}

\bibitem{sum} C.H. Jin, Mod. Phys. Lett. A 14 (1999) 1163;
Phys. Rev. D 62 (2000) 014020.

\bibitem{LC} C.H. Jin, E.A. Paschos, in {\it Proceedings of the
International
Symposium on Heavy Flavor and Electroweak Theory}, Beijing, China,
1995, edited by C.H. Chang and C.S. Huang (World Scientific, Singapore,
1996), p.~132; hep-ph/9504375;
C.H. Jin, Phys. Rev. D 56 (1997) 2928;
C.H. Jin, E.A. Paschos, Eur. Phys. J. C 1 (1998) 523;
C.H. Jin, Phys. Rev. D 56 (1997) 7267;
Eur. Phys. J. C 11 (1999) 335.

\bibitem{isgur} N. Isgur, M.B. Wise, Phys. Lett. B 232 (1989) 113;
B 237 (1990) 527.

\bibitem{eichten} E. Eichten, B. Hill, Phys. Lett. B 234 (1990) 511;
B 243 (1990) 427.

\bibitem{georgi} H. Georgi, Phys. Lett. B 240 (1990) 447.

\bibitem{jaffe} R.L. Jaffe, X. Ji, Nucl. Phys. B 375 (1992) 527.

\bibitem{ball} P. Ball, V.M. Braun, Phys. Rev. D 49 (1994) 2472.

\bibitem{latt} A.S. Kronfeld, J.N. Simone,
Phys. Lett. B 490 (2000) 228.

\bibitem{PDG} Particle Data Group, D.E. Groom et al.,
Eur. Phys. J. C 15 (2000) 1.
\end{references}
\end{document}